\documentclass{article}

\PassOptionsToPackage{numbers, compress, sort}{natbib}
\usepackage[preprint]{neurips_2025}

\usepackage[utf8]{inputenc} 
\usepackage[T1]{fontenc}    
\usepackage{hyperref}       
\usepackage{url}            
\usepackage{booktabs}       
\usepackage{amsfonts}       
\usepackage{nicefrac}       
\usepackage{microtype}      
\usepackage{xcolor}         

\usepackage{graphicx}
\usepackage{amsmath,amssymb,gensymb}
\usepackage{siunitx}
\usepackage{textcomp,marvosym}
\DeclareSIUnit{\Molar}{M}

\usepackage[aboveskip=1pt,labelfont=bf,labelsep=period,justification=raggedright,singlelinecheck=off]{caption}

\title{Design, Simulation, and Fabrication of a Hexagonal Microfluidic Platform for Culturing Neurons}

\author{%
  Maxx Yung \\
  Department of Materials Science and Engineering\\
  University of Pennsylvania\\
}

\begin{document}

\maketitle

\begin{abstract}
  Developing an organoid computing platform from neurons in vitro demands stable, precisely controlled microenvironments. To address this requirement, we designed, simulated, and fabricated a microfluidic device featuring hexagonal wells ($34.64\:\mu\text{m}$ side length) in a honeycomb array connected by $20\:\mu\text{m}$ channels. Computational fluid dynamics (CFD) modeling, validated by high mesh quality ($0.934$ orthogonal quality) and robust convergence, confirmed the architecture supports flow regimes ideal for ensuring cell viability. At target flow rates of $0.1-1\:\mu\text{L/min}$, simulations revealed the extrapolated pressure differential across the full $50,000\:\mu\text{m}$ device remains within stable operating limits at $177\:\text{kPa}$ (average) and $329\:\text{kPa}$ (maximum). Photolithography successfully produced this architecture, with only minor corner rounding observed at feature interfaces. This work therefore establishes a computationally validated and fabricated platform, paving the way for experimental flow characterization and subsequent neural integration.
\end{abstract}

\section{Introduction}
The rapid growth in artificial intelligence computational demands has exposed fundamental limitations in traditional computing technologies \cite{xuLargescalePhotonicChiplet2024}. As Moore's law approaches its physical limits, computing systems face increasing energy efficiency challenges \cite{muralidharEnergyEfficientComputing2022}, prompting companies to explore extreme solutions like nuclear-powered data centers. Digital computing architectures struggle to match the remarkable energy efficiency observed in biological neural networks, which achieve complex computations through probabilistic and memory-efficient computation \cite{houshmandBenchmarkingModelingAnalog2023}.

Biological computing using living neuronal networks presents a promising alternative to conventional computational approaches \cite{smirnovaBiocomputingOrganoidIntelligence2024,khajehnejadBiologicalNeuronsCompete2024, caiBrainOrganoidReservoir2023, kaganVitroNeuronsLearn2022}. However, the success of such systems critically depends on maintaining precise cellular environments. Critical parameters include maintaining pH at 7.2-7.4 through continuous $CO_2$ buffering, temperature stability at $37\pm0.5\degree\text{C}$, and more importantly, regular media exchange every $48-72$ hours to replenish glucose nutrient media while removing metabolic waste products \cite{kaganVitroNeuronsLearn2022, seilerModularAutomatedMicrofluidic2022}. Manual media exchange, typically performed every 1–3 days, causes erratic nutrient fluctuations, toxic buildup, and exposure to suboptimal conditions during media replenishment. Automated microfluidic systems address these issues by maintaining precise, continuous nutrient delivery and waste removal, minimizing cellular stress and replicating physiological conditions more accurately \cite{seilerModularAutomatedMicrofluidic2022}.

This undergraduate research project aims to simulate and develop a microfluidic platform optimized for future organoid computing applications. The focus will be on understanding computational fluid dynamics, lithography techniques, and designing a microfluidic system to achieve precise flow control ($0.1-1~\text{ \micro L/min}$).

Although the integration of microelectrode arrays and neural cultures was not researched here, this project design will account for future integration requirements based on established parameters in the field and will serve as preparatory research for future projects.

\section{Methodology}
\subsection{Computational Fluid Dynamics}
\begin{figure}[!h]
  \centering
  \includegraphics[width=0.48\textwidth]{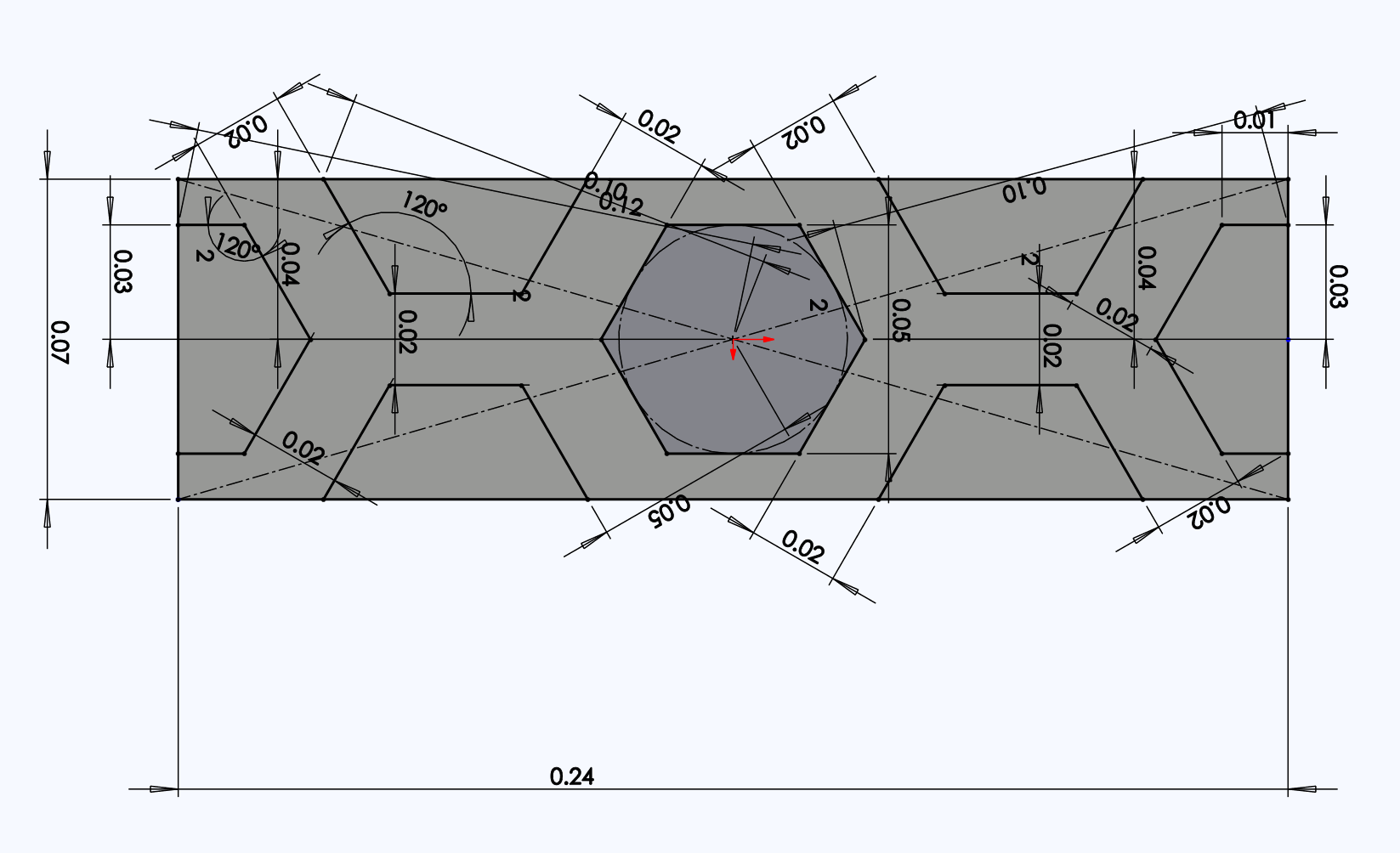}
  \includegraphics[width=0.48\textwidth]{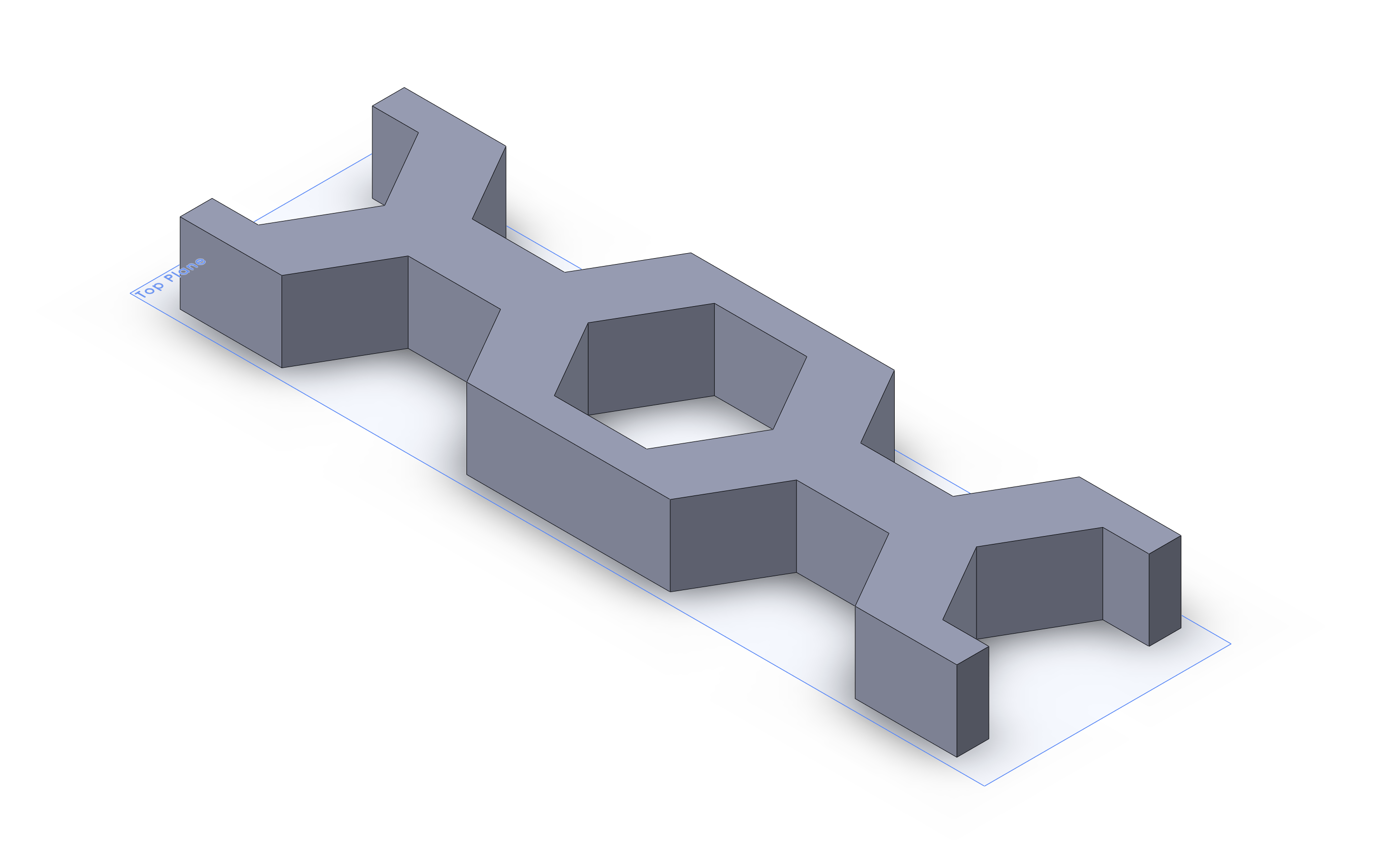}
  \caption{{\bf SolidWorks Channel Design.}
  A: Sketch relations used to define the microfluidic channel dimensions. B: 3D representation of the single channel geometry with 0.2424mm height.}
  \label{fig:cad}
\end{figure}

Finite element analysis was performed on a simulated single microfluidic channel. The modular design allows single-channel flow analysis, saving computational resources and enabling extrapolation to the full device. The geometry was created as a CAD mockup in SolidWorks (Fig. \ref{fig:cad}).

\begin{figure}[!h]
  \centering
  \includegraphics[width=1\textwidth]{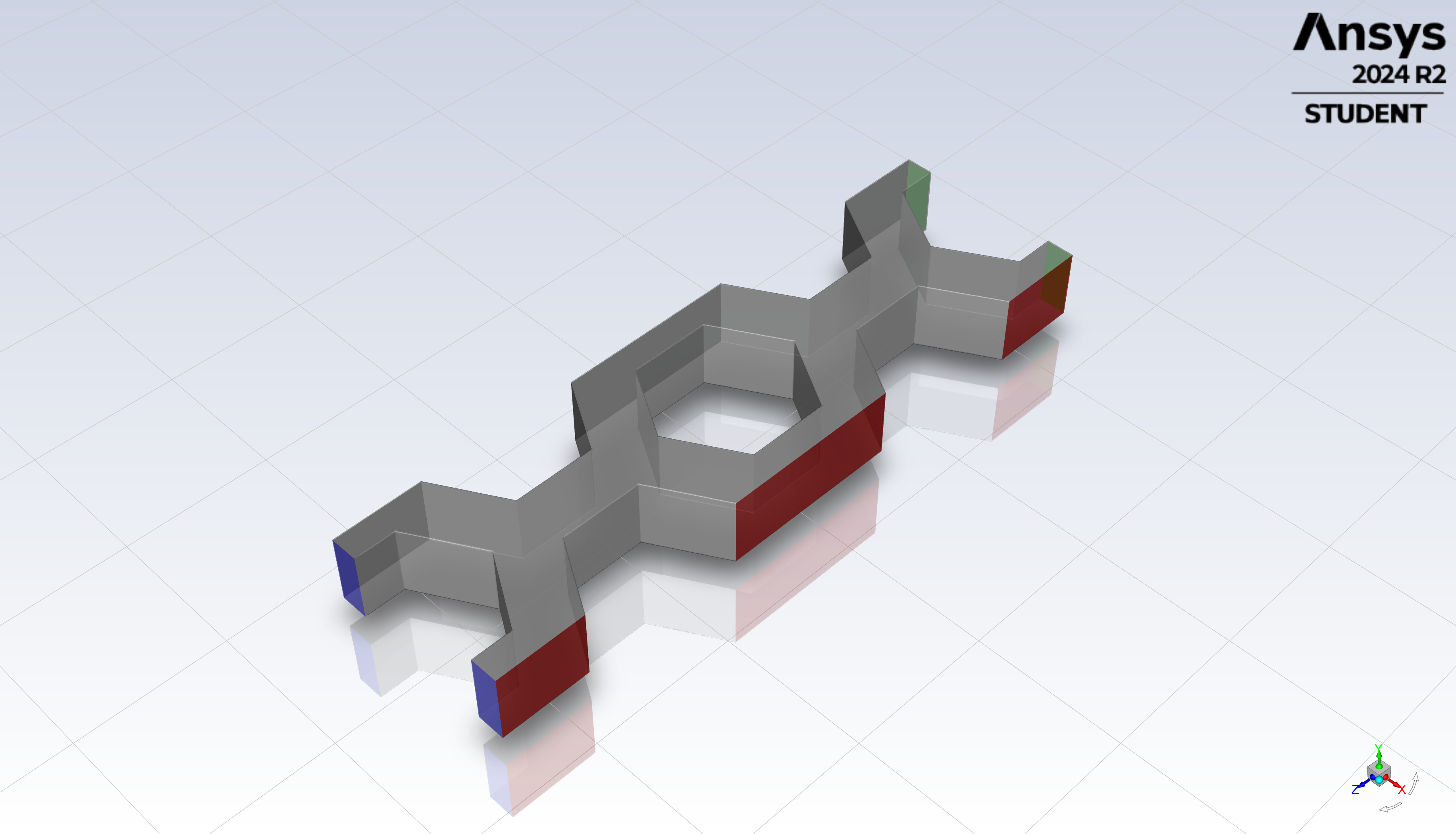}
  \caption{{\bf ANSYS Fluent Geometry Import.}
  Imported channel geometry showing boundary condition definitions of the channel walls and inlet faces for fluid simulation.}
  \label{fig:ansys_geom}
\end{figure}

The completed geometry was imported into ANSYS Fluent (Fig. \ref{fig:ansys_geom}), where input and output faces were defined (highlighted in blue and green, respectively). Symmetry walls were defined on the left and right sides of the channel (highlighted in red).

\begin{figure}[!h]
  \centering
  \includegraphics[width=1\textwidth]{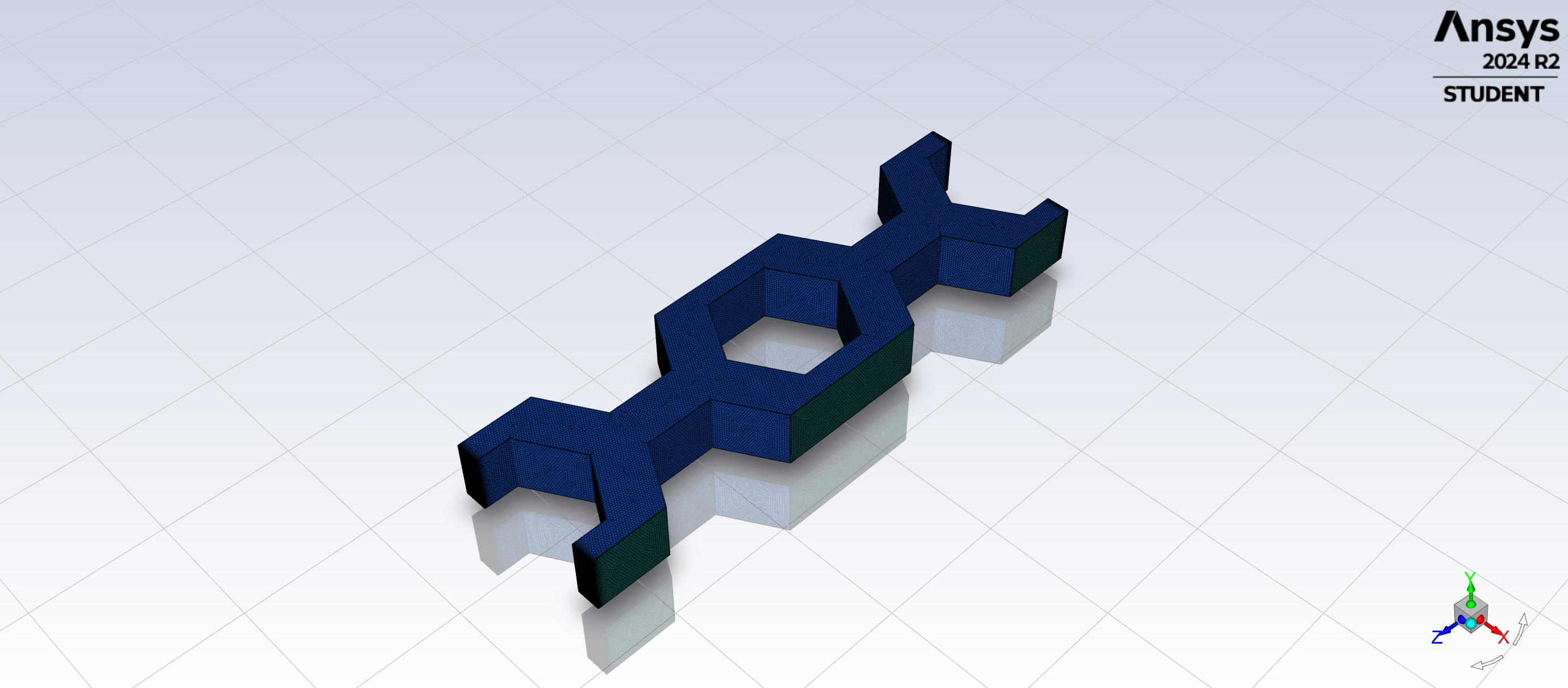}
  \caption{{\bf Complete Mesh Overview.}
  Full mesh visualization of overall mesh quality across the channel geometry.}
  \label{fig:ansys_mesh}
\end{figure}

\begin{figure}[!h]
  \centering
  \includegraphics[width=0.48\textwidth]{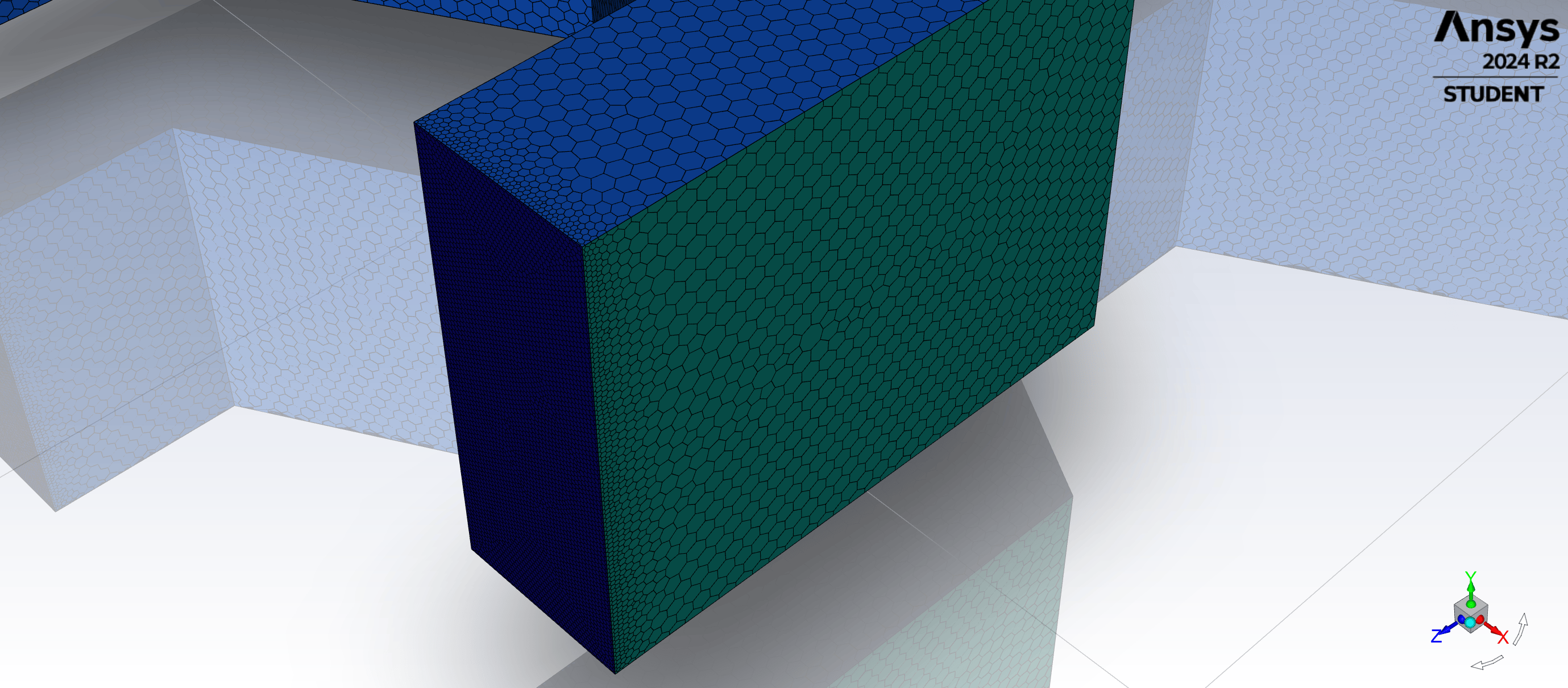}
  \includegraphics[width=0.48\textwidth]{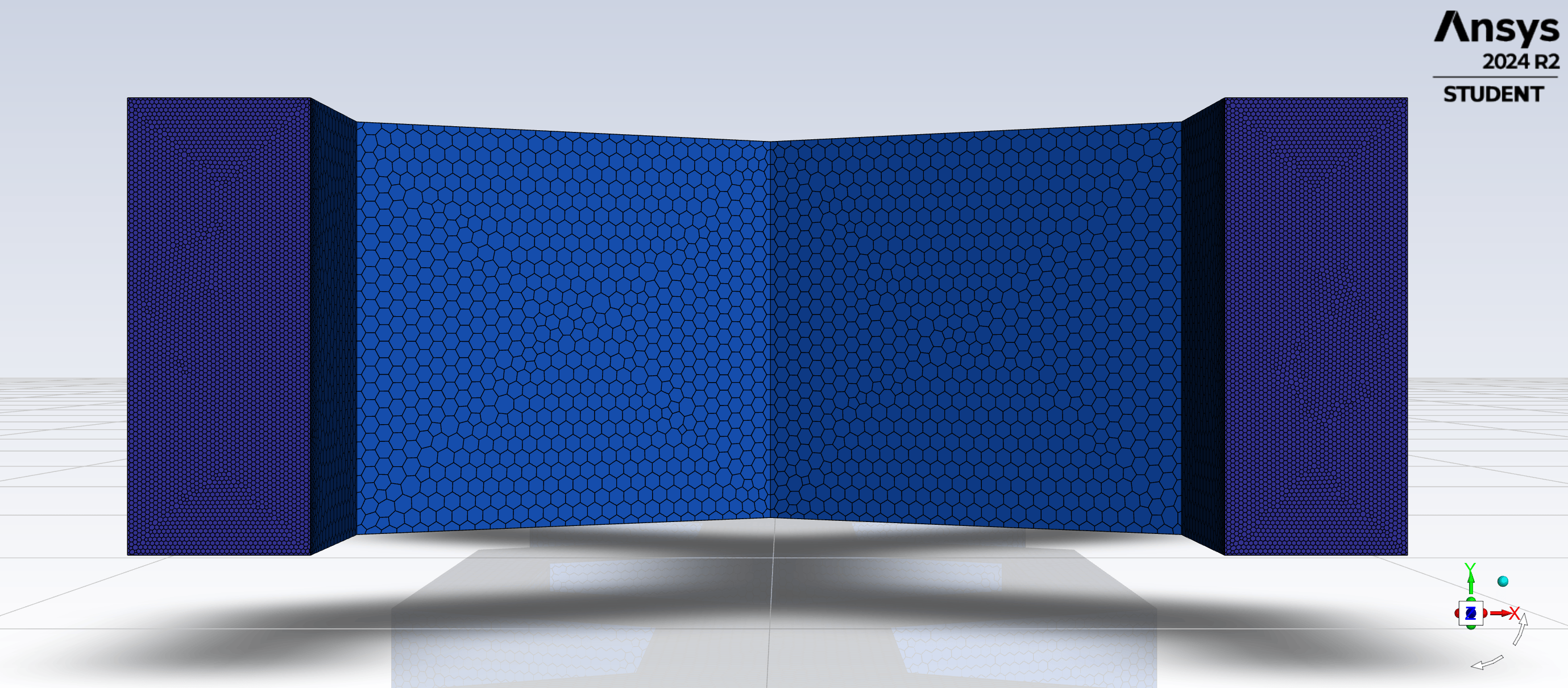}
  \caption{{\bf Detailed Mesh Views.}
  A: Detailed view of mesh refinement at inlet region. B: Front view displaying mesh transition.}
  \label{fig:mesh_detail}
\end{figure}

The complete mesh structure (Fig. \ref{fig:ansys_mesh}) was configured using double precision settings, with local sizing added specifically at the inlet and outlet faces to enhance precision (Fig. \ref{fig:mesh_detail}). These boundary regions were configured with a growth rate of $1.2$ and a target mesh size of $0.25~\mu\text{m}$. Surface mesh generation parameters were set with a minimum size of $0.2368038~\mu\text{m}$ and a maximum size of $1~\mu\text{m}$. The geometry consisted solely of fluid regions with no void spaces, and fluid boundary types were set to wall conditions by default with no-slip conditions applied. Boundary conditions specified a velocity inlet for the inlet face ($0.0002~\text{m}/\text{s}$, $0.0005~\text{m}/\text{s}$, and $0.001~\text{m}/\text{s}$) and a pressure outlet for the outlet face. The two lateral channel walls were designated as symmetry boundaries to simulate the behavior of multiple parallel channels, with the entire region labeled as a fluid domain. Temperature effects were not considered in the simulation, with water modeled using its standard properties at room temperature.

\begin{figure}[!h]
  \centering
  \includegraphics[width=0.48\textwidth]{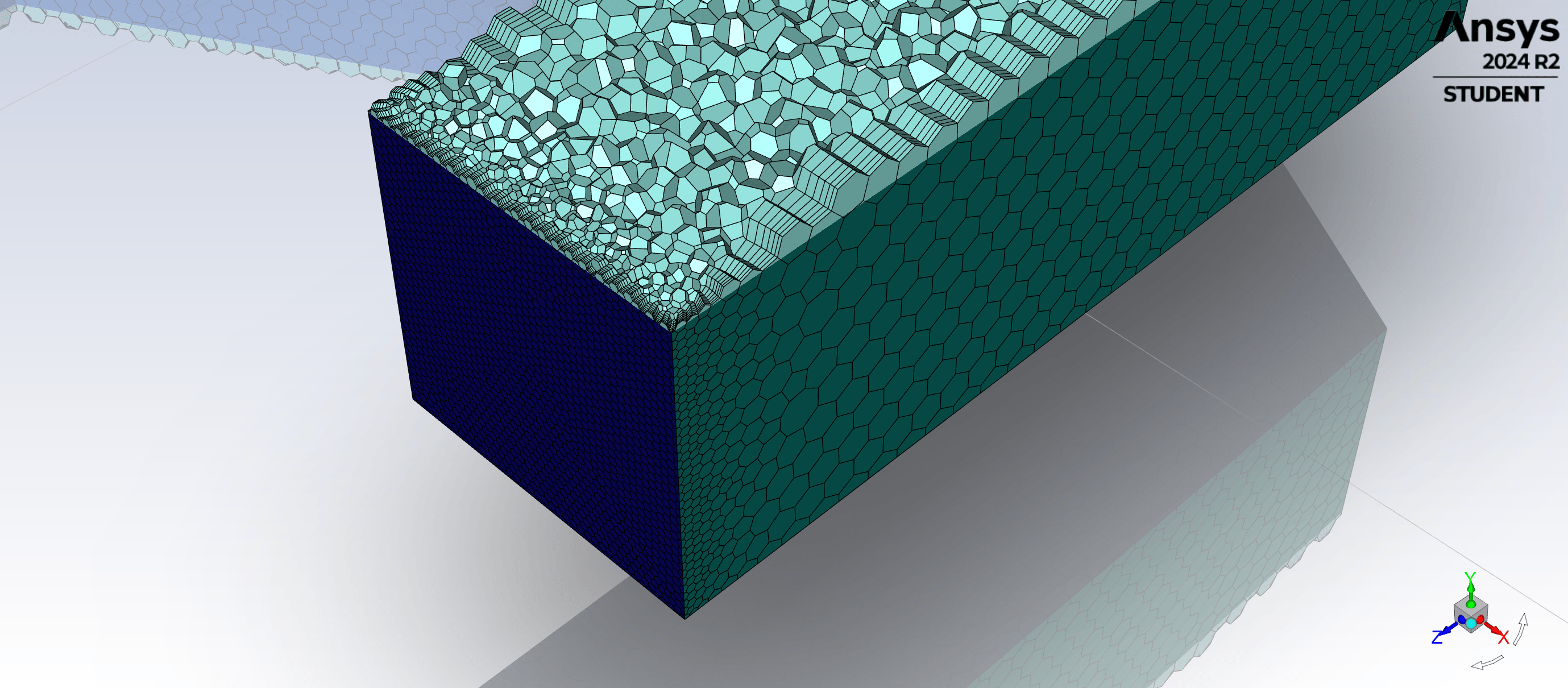}
  \includegraphics[width=0.48\textwidth]{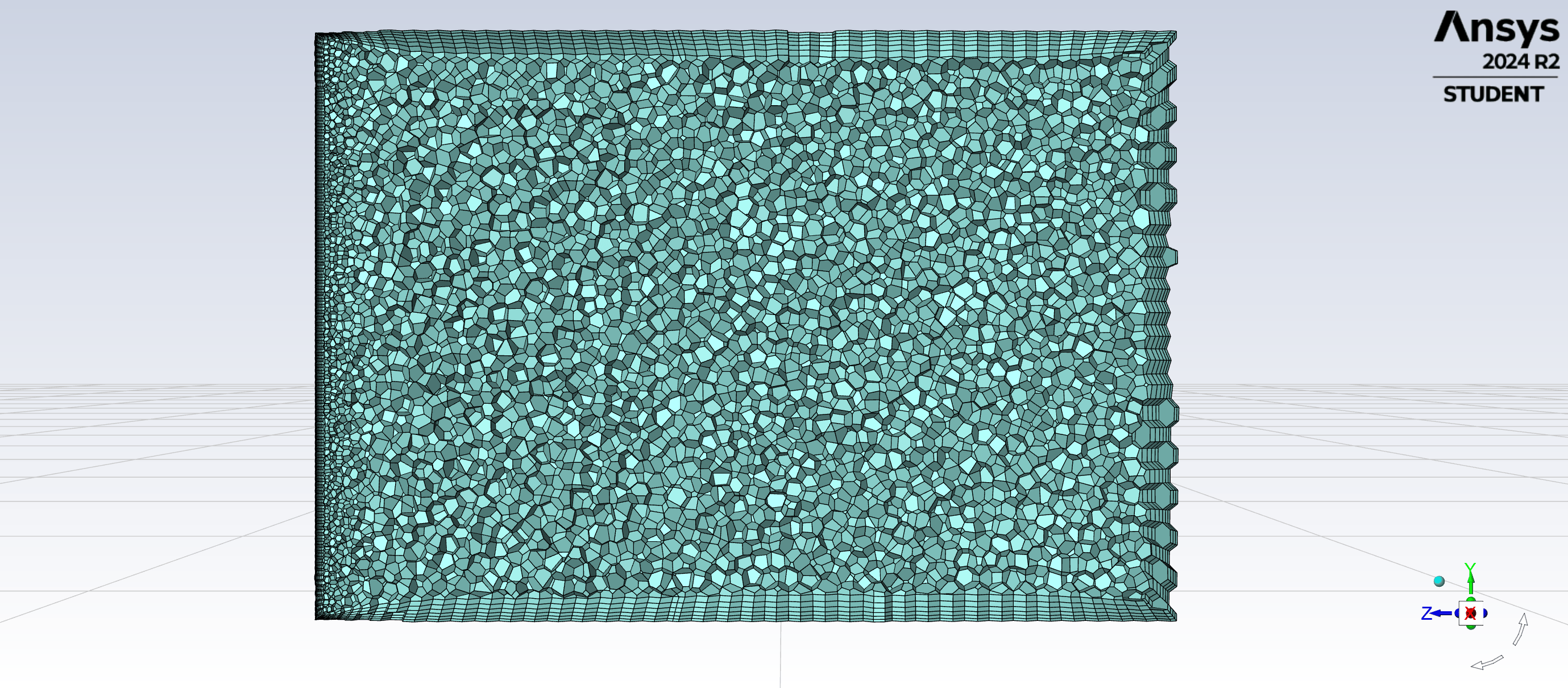}
  \caption{{\bf Individual Cell Visualization.}
  A: Close-up view of polyhedral mesh elements near inlet region. B: Detailed visualization of individual cell quality in critical flow areas.}
  \label{fig:cell_detail}
\end{figure}

A single boundary layer was implemented with default settings, comprising $6$ layers, a transition ratio of $0.272$, and a growth rate of $1.2$ across all fluid regions and zones. Individual cell quality can be observed in the critical regions (Fig. \ref{fig:cell_detail}), where the solver utilized polyhedral mesh elements with a growth rate of $1.2$ and maximum cell length of $0.8~\mu\text{m}$. Only laminar flow was modeled. The simulation used aluminum as the solid material and water as the fluid medium designated throughout the entire domain. The simulation was executed for $1000$ iterations using the SIMPLEC solution scheme.

\subsection{Fabrication}

\begin{figure}[!h]
  \centering
  \includegraphics[width=1\textwidth]{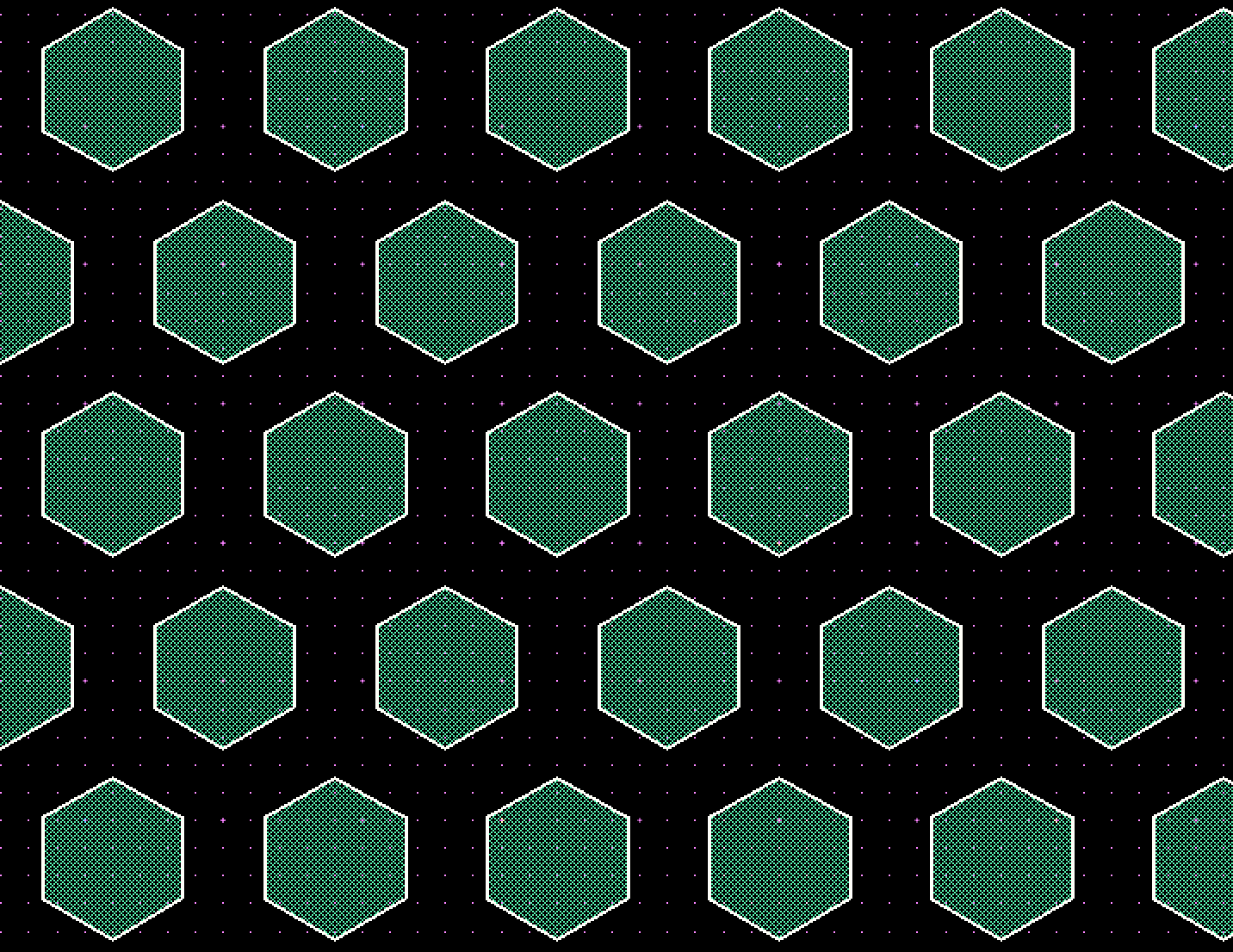}
  \caption{{\bf Photolithography Mask Layout.}
  Final mask design showing the hexagonal well array pattern created in Layout Editor.}
  \label{fig:layout}
\end{figure}

The microfluidic channel design was first created in SolidWorks to establish the dimensional  relationships of the device (Fig.\ref{fig:cad}). This CAD design was then translated into a mask layout using LayoutEditor software. The final mask design consists of an array of regular hexagonal wells with side lengths of $34.64~\mu\text{m}$ arranged in a honeycomb pattern (Fig.\ref{fig:layout}). Microfluidic channels connecting each well were designed with widths of $20~\mu\text{m}$. The complete layout was exported in GDSII format for direct-write lithography.

\begin{figure}[!h]
  \centering
  \includegraphics[width=1\textwidth]{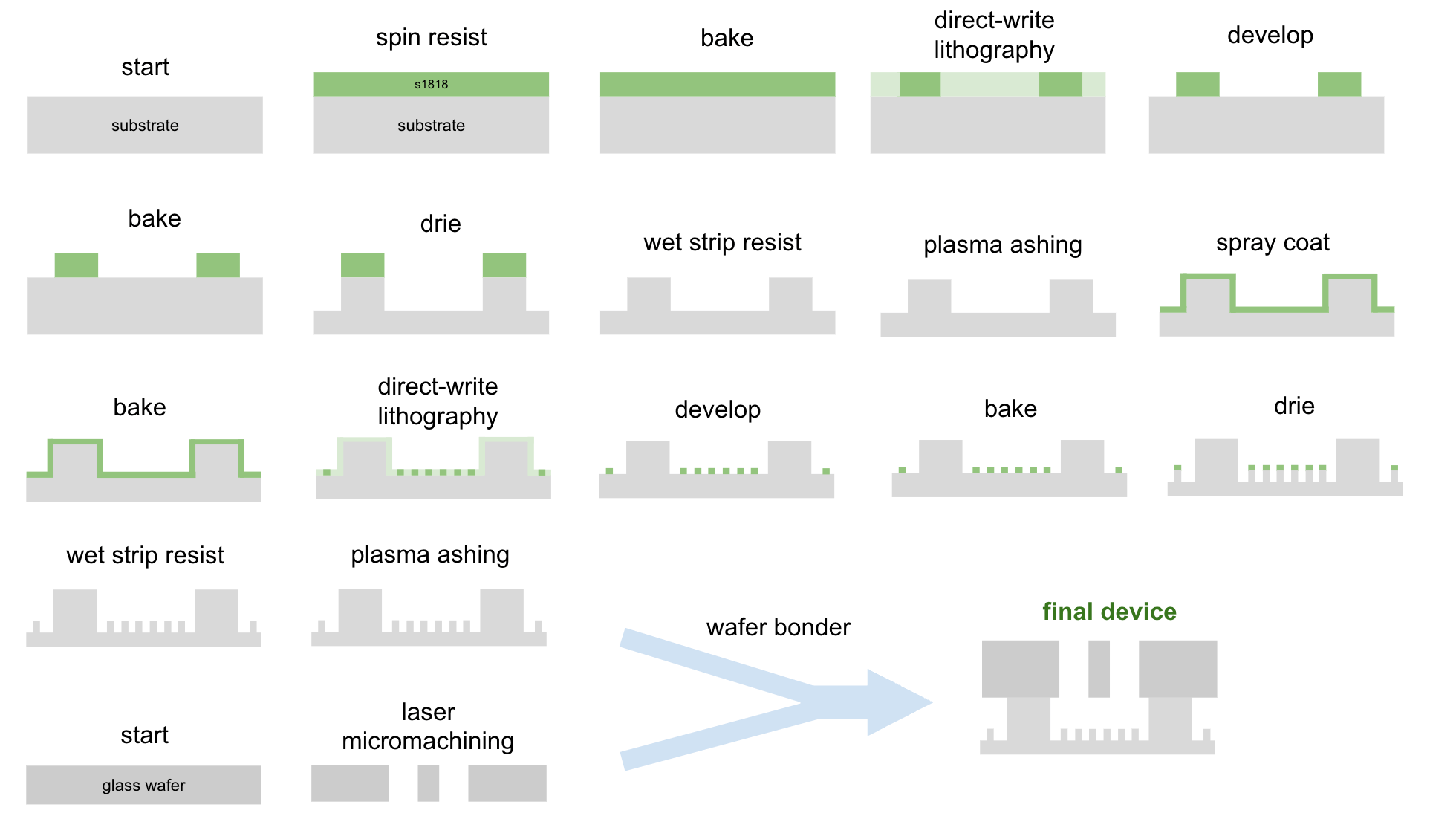}
  \caption{{\bf Microfluidic Device Fabrication Process Flow.}
  Schematic illustration of the complete fabrication process to create the enclosed microfluidic device.}
  \label{fig:fabrication}
\end{figure}

The microfluidic device was fabricated using standard photolithography techniques. A silicon wafer was cleaned with acetone, isopropyl alcohol, and deionized water before spin-coating with SU-8 1818 photoresist. The photoresist was spin-coated and soft-baked to remove excess solvent. The substrate was then exposed to UV light through direct-write lithography to define the microfluidic channel patterns. After exposure, a post-exposure bake, photoresist development, and a hard bake step were done to strengthen the patterned features. Final steps involved plasma ashing and a wet strip resist process to ensure cleaner feature definition. Finally, the device was sealed using a wafer bonding process to create the enclosed microfluidic channels, resulting in the final functional device. A detailed step-by-step fabrication process flow is shown in Fig.~\ref{fig:fabrication}.

\begin{figure}[!h]
  \centering
  \includegraphics[width=1\textwidth]{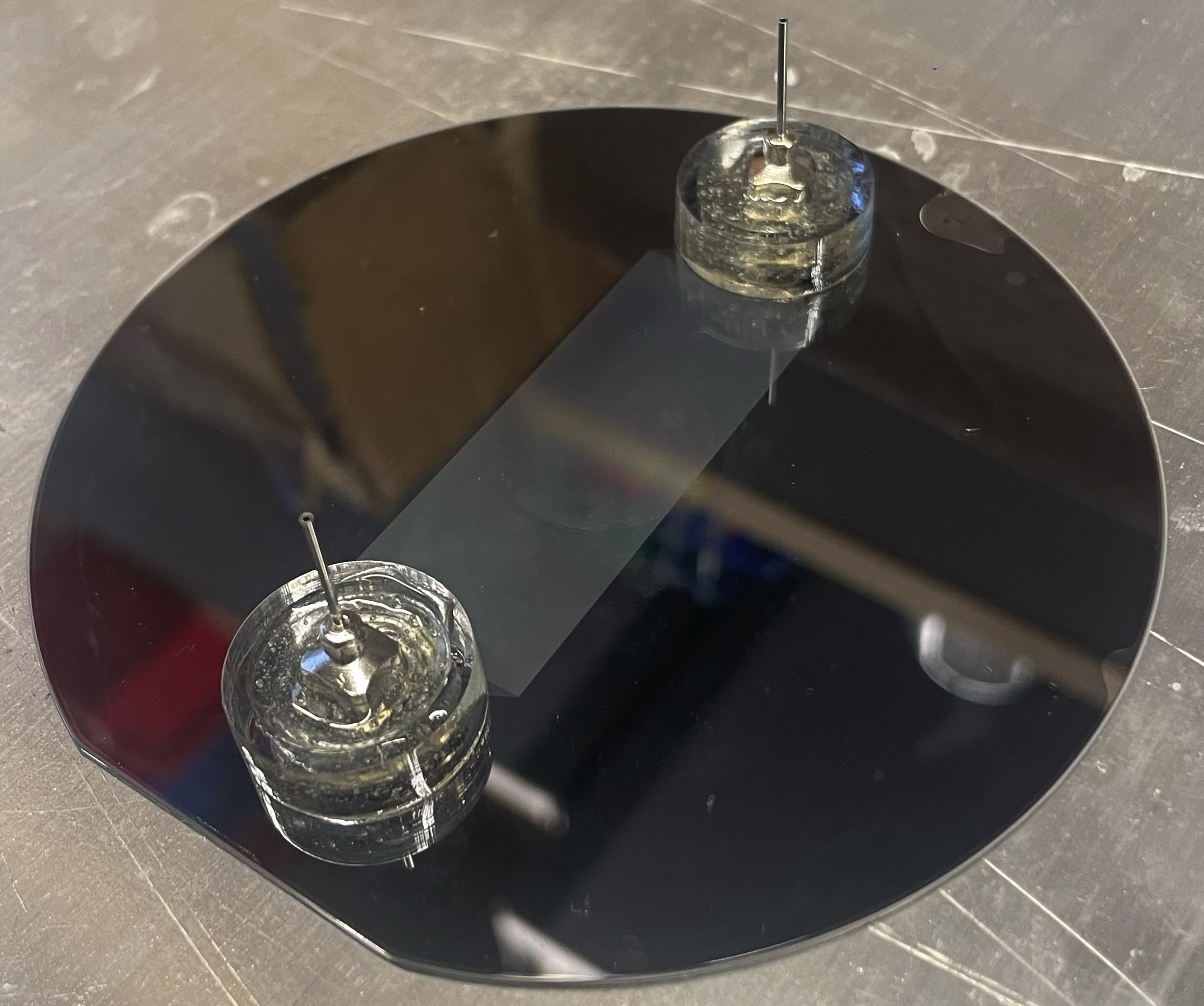}
  \caption{{\bf Microfluidic Device Testing Setup.}
  The microfluidic device with two connecting ports for flow testing.}
  \label{fig:testing}
\end{figure}

To measure the performance of the fabricated microfluidic device, a testing protocol was designed. The proposed protocol would connect the device's inlet and outlet ports to a pressure-driven flow system using UV-curable adhesive to create sealed connections, as shown in Fig.~\ref{fig:testing}. The testing protocol would be performed at specified flow rates ranging from $0.1-1~\text{\micro L/min}$. Pressure measurements would be recorded at both inlet and outlet ports to calculate the pressure drop across the device.

\section{Results}

\subsection{Mesh Quality and Convergence Analysis}

The final mesh optimization achieved 988,203 cells with over 3.8 million interior nodes, approaching the 1,000,000 computational limitations for ANSYS Student Edition while maintaining sufficient resolution for accurate flow field calculation. Mesh quality metrics confirmed the simulation reliability with an average orthogonal quality of 0.934 (minimum 0.317), average skewness of 0.008 (maximum 0.257), and maximum aspect ratio of 25.48. Residual convergence thresholds of $1\times10^{-3}$ for continuity and $1\times10^{-5}$ for velocity components were achieved.

The static pressure profile reveals two distinct pressure drop characteristics across the simulated channel segment. The average pressure decreases from approximately $850~\text{Pa}$ to $0~\text{Pa}$, while maximum pressures peak at $1.58~\text{kPa}$ before dropping to 0 Pa at the outlet. At flow rates of $0.1$--$1\:\mu\text{L/min}$, pressure analysis revealed drops from $850\:\text{Pa}$ to $0\:\text{Pa}$ (average) and $1.58\:\text{kPa}$ peak pressures across the simulated segment. Extrapolated to the full $50,000\:\mu\text{m}$ device length, this indicates pressure differentials of $177\:\text{kPa}$ (average) and $329\:\text{kPa}$ (maximum). Photolithographic fabrication successfully produced the designed patterns, though, with some edge roughness and corner rounding at feature interfaces. Future work will involve experimental flow characterization to validate computational predictions and optimize device performance.

The pressure drop across the full device length was extrapolated through an assumed linear relationship under the Hagen-Poiseuille equation:

\begin{equation}
\Delta P_{total} = \Delta P_{unit} \times \frac{L_{total}}{L_{unit}}
\end{equation}

where $\Delta P_{unit}$ is the pressure drop across one unit, $L_{unit}$ is the unit length, and $L_{total}$ is the total device length.

For the average pressure drop:
\begin{equation}
\Delta P_{total} = 850\:\text{Pa} \times \frac{50,000\:\mu\text{m}}{240\:\mu\text{m}} = 177,083\:\text{Pa} \approx 177\:\text{kPa}
\end{equation}

For the maximum pressure drop:
\begin{equation}
\Delta P_{total} = 1,580\:\text{Pa} \times \frac{50,000\:\mu\text{m}}{240\:\mu\text{m}} = 329,167\:\text{Pa} \approx 329\:\text{kPa}
\end{equation}

\begin{figure}[!h]
  \centering
  \includegraphics[width=1\textwidth]{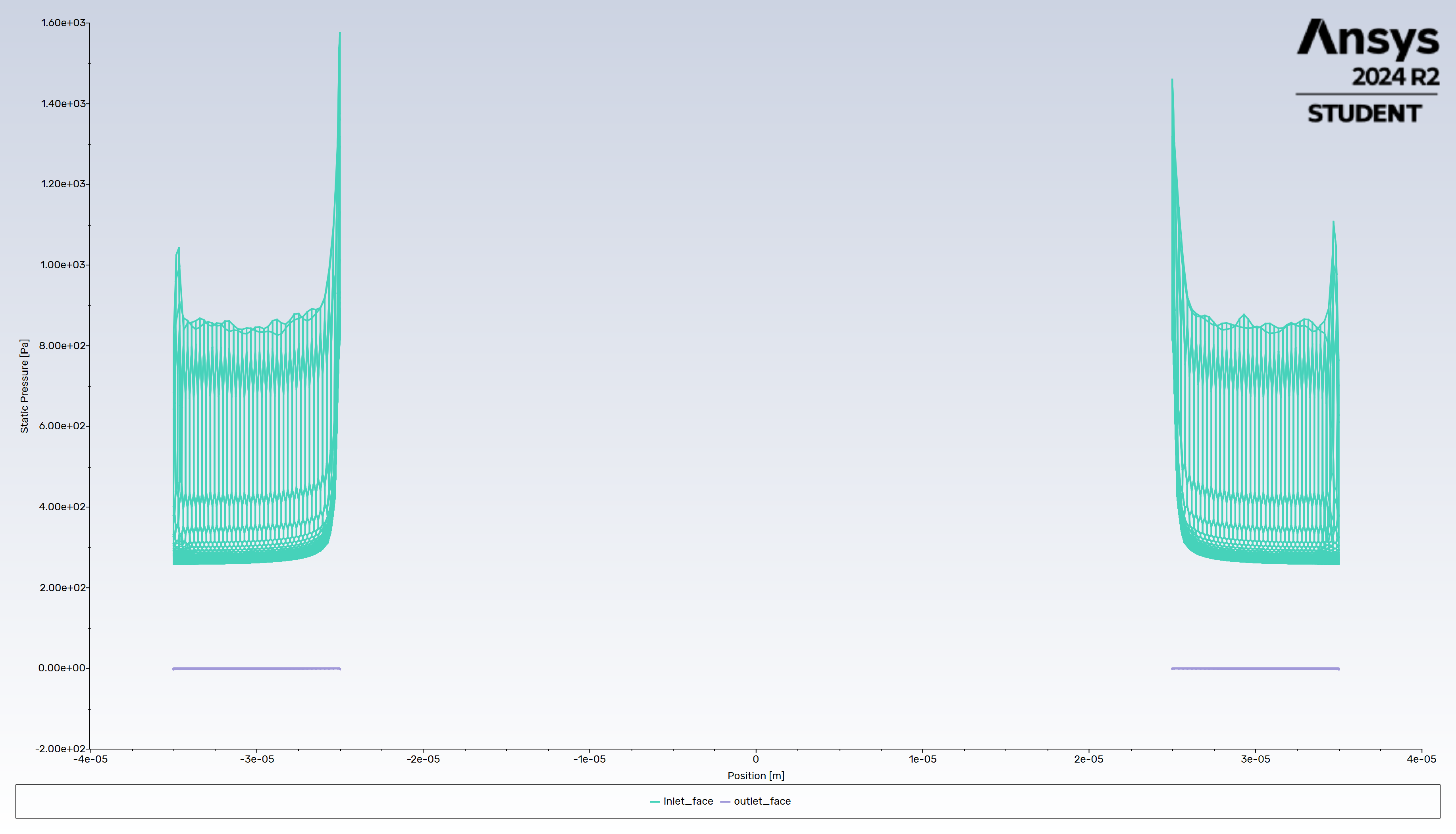}
  \caption{{\bf Static Pressure Distribution.}
  Static pressure profile showing both average $(850~\text{Pa})$ and peak ($1.58~\text{kPa}$) pressure drops across the channel segment.}
  \label{fig:static_pressure}
\end{figure}

\begin{figure}[!h]
  \centering
  \includegraphics[width=1\textwidth]{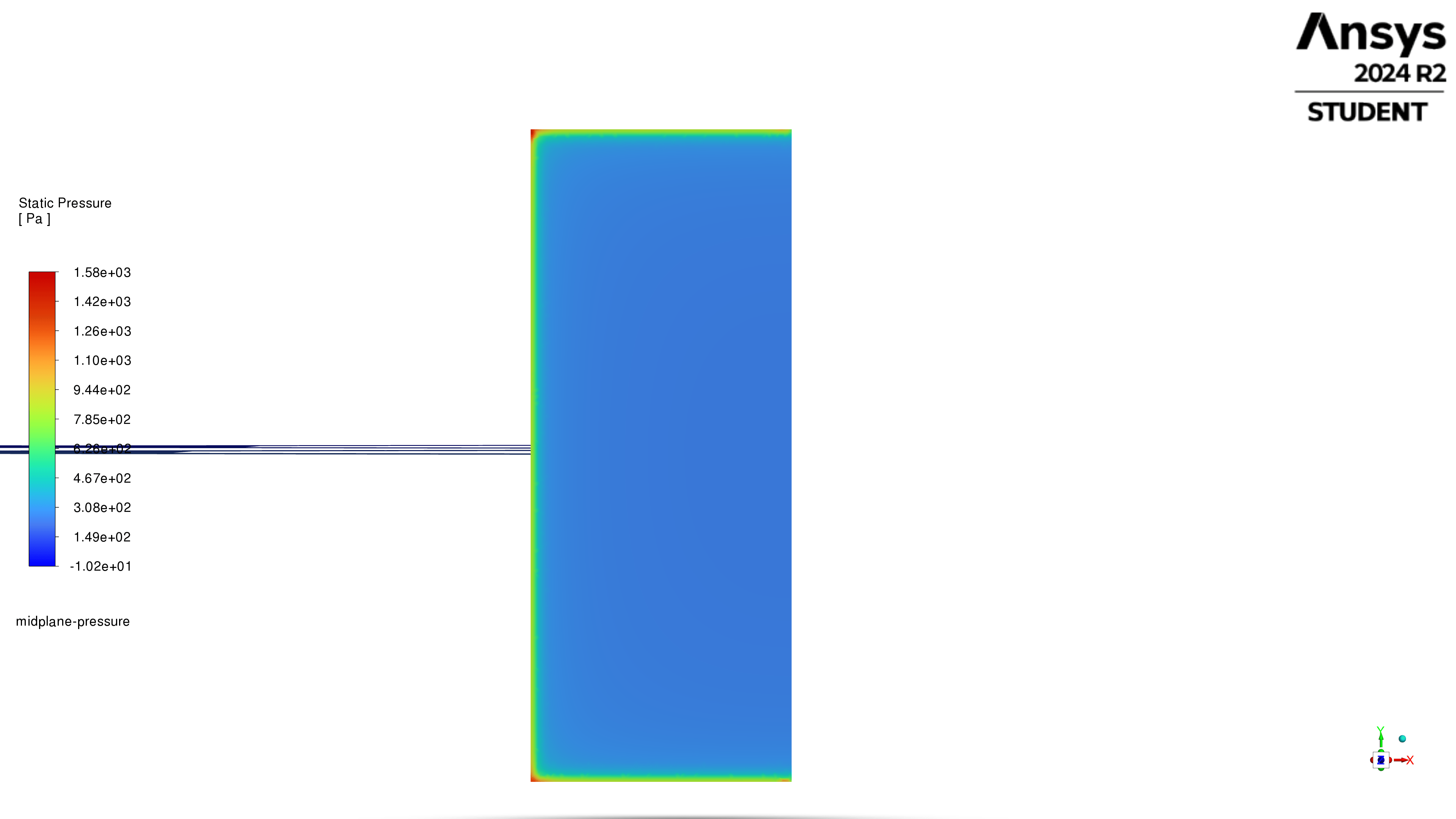}
  \caption{{\bf Static Pressure Distribution at Inlet.}
  Static pressure profile for the inlet, with uniform pressure region and characteristic boundary gradients near the wall.}
  \label{fig:static_pressure_inlet}
\end{figure}

The static pressure distribution at the inlet face of the microfluidic channel reveals a notable gradient, with an average pressure of \(3.08 \times 10^2~\text{Pa}\) and localized peaks reaching \(1.58 \times 10^3~\text{Pa}\) at the corners as shown in Fig.~\ref{fig:static_pressure_inlet}. This pressure concentration at the corners is characteristic of fluid flow and arises from abrupt changes in flow direction and velocity gradients. At the sharp corners, the fluid velocity decreases locally, leading to a corresponding rise in static pressure as described by Bernoulli's principle. In contrast, regions along the central inlet face experience more uniform flow, resulting in lower pressures.

\begin{figure}[!h]
  \centering
  \includegraphics[width=1\textwidth]{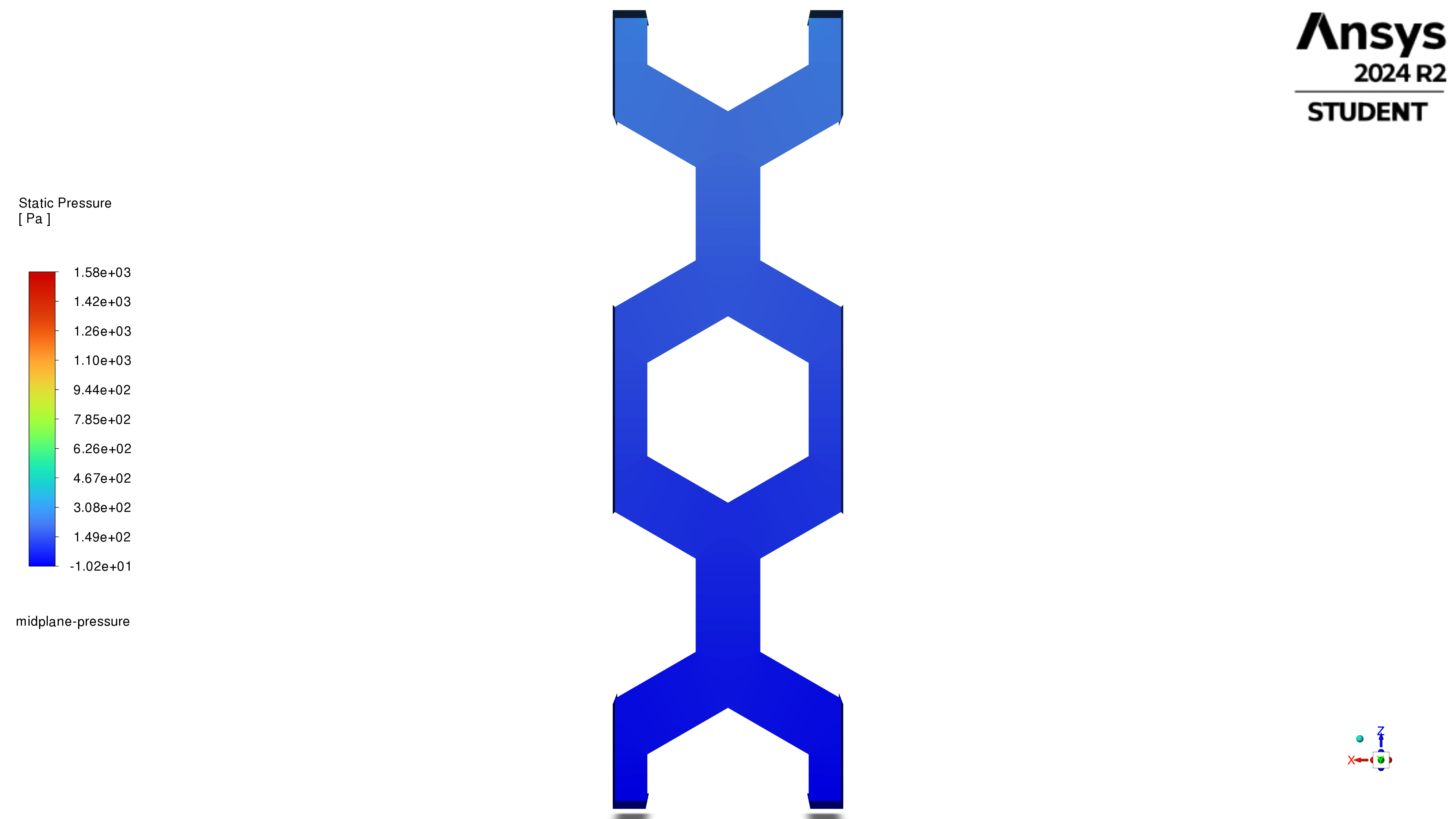}
  \caption{{\bf Static Pressure Distribution Across Channel.}
  Static pressure profile illustrating the pressure drop across the length of a single microfluidic channel. The design’s symmetry shows uniform flow distribution, with minimal pressure variations.}
  \label{fig:static_pressure_total}
\end{figure}

The static pressure distribution across the microfluidic channel exhibits a characteristic gradient from inlet to outlet, as shown in Fig.~\ref{fig:static_pressure_total} and detailed in Table~\ref{tab:pressure_distribution}. The pressure measurements demonstrate a well-designed initial geometry that minimizes pressure losses while maintaining stable flow conditions.

\begin{figure}[!h]
  \centering
  \includegraphics[width=0.48\textwidth]{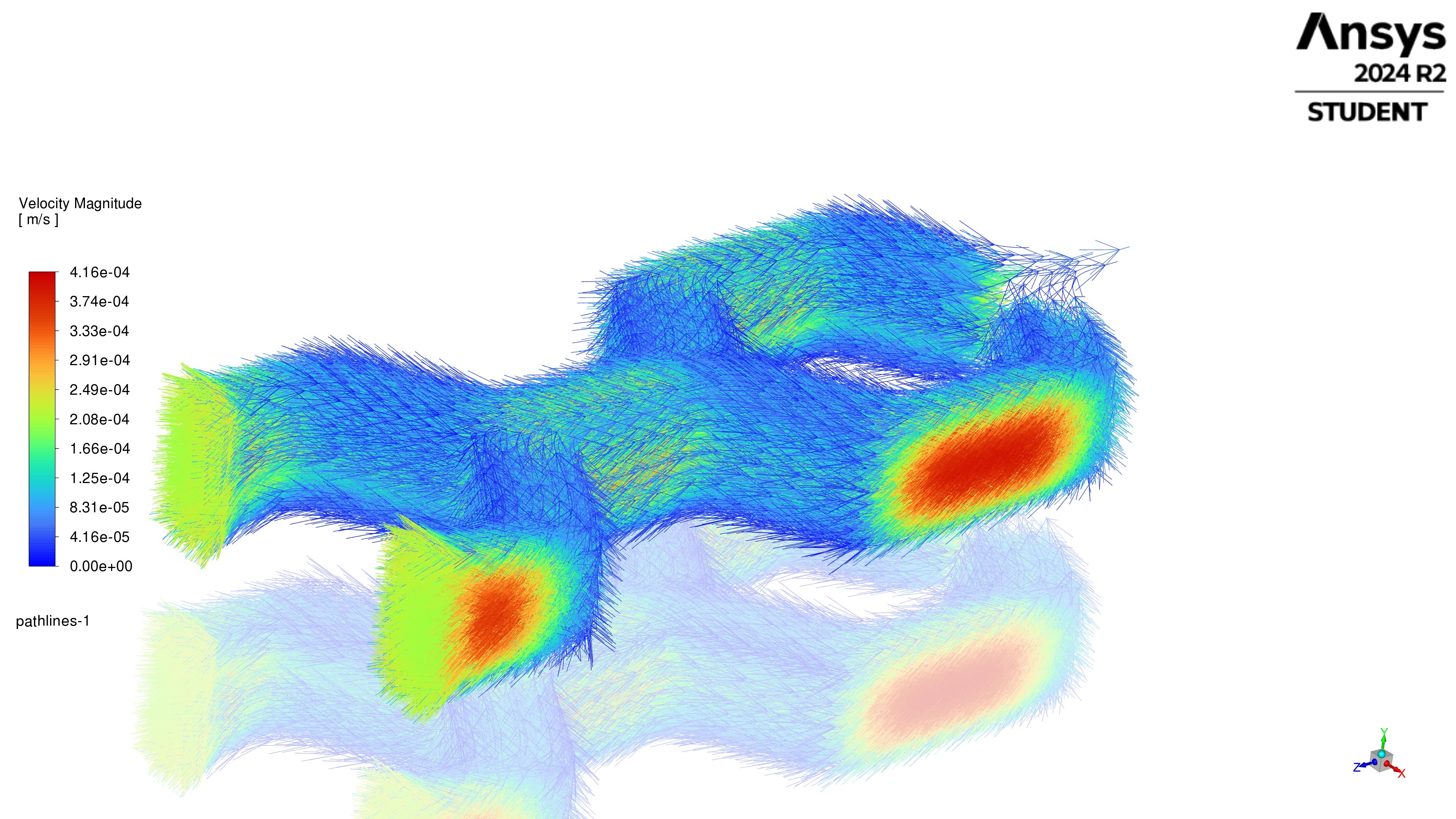}
  \includegraphics[width=0.48\textwidth]{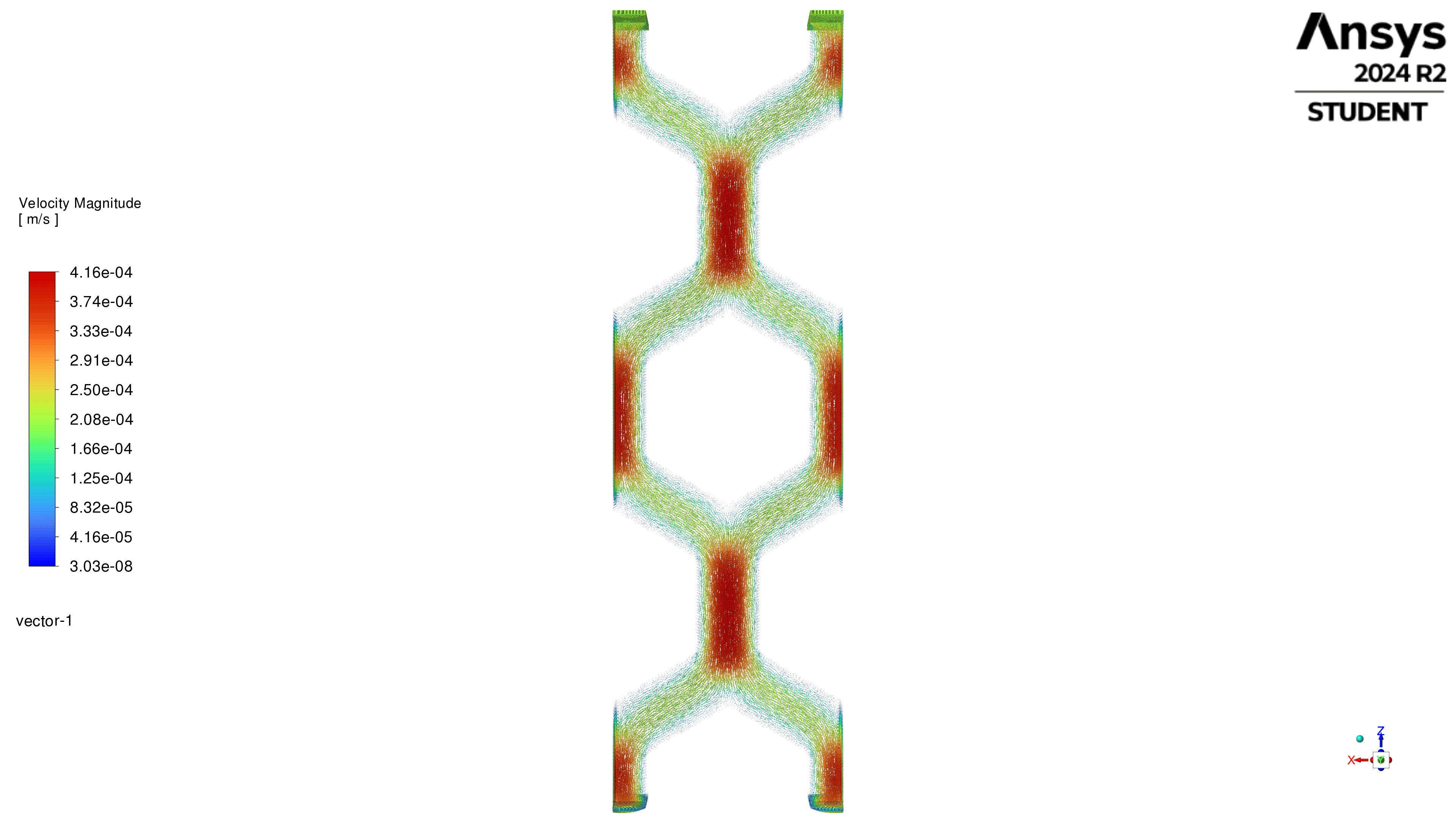}
  \caption{{\bf Velocity Distribution Across Channel.}
  Velocity profile showing fluid flow pathlines (left) and top view (right) across the microfluidic channel. The pathlines illustrate the flow direction and velocity magnitude, with higher velocities (red) in the center and lower velocities (blue/green) near the walls.}
  \label{fig:velocity}
\end{figure}

The velocity profile was found to be higher near the inlet and outlet channels and symmetry walls ($4.16 \times 10^{-4}$ m/s) and lower in the main channel ($4 \times 10^{-5}$ m/s), as shown in Fig.~\ref{fig:velocity}.

All quantitative measurements and analysis of the pressure distribution, including inlet and outlet pressures, pressure gradients, and velocity measurements, are summarized in Table~\ref{tab:pressure_distribution} below. Additional results from simulating inlet velocity at $0.0005~\text{m/s}$ and $0.001~\text{m/s}$ are also included in the table.

The volumetric flow rate $Q$ was calculated using the equation:

\begin{equation}
Q = A \times V
\end{equation}

where $A$ is the cross-sectional area of the inlet and $V$ is the fluid velocity at the inlet. The inlet has dimensions of $0.025~\text{mm} \times 0.04~\text{mm}$, giving a cross-sectional area:

\begin{equation}
A = 0.025~\text{mm} \times 0.04~\text{mm} = 0.001~\text{mm}^2 = 1 \times 10^{-9}~\text{m}^2
\end{equation}

For each inlet velocity, the volumetric flow rate was calculated as follows:

For $V = 0.0002~\text{m/s}$:
\begin{equation}
Q = (1 \times 10^{-9}~\text{m}^2)(0.0002~\text{m/s}) = 2 \times 10^{-13}~\text{m}^3/\text{s} = 0.012~\mu\text{L}/\text{min}
\end{equation}

For $V = 0.0005~\text{m/s}$:
\begin{equation}
Q = (1 \times 10^{-9}~\text{m}^2)(0.0005~\text{m/s}) = 5 \times 10^{-13}~\text{m}^3/\text{s} = 0.03~\mu\text{L}/\text{min}
\end{equation}

For $V = 0.001~\text{m/s}$:
\begin{equation}
Q = (1 \times 10^{-9}~\text{m}^2)(0.001~\text{m/s}) = 1 \times 10^{-12}~\text{m}^3/\text{s} = 0.06~\mu\text{L}/\text{min}
\end{equation}

\begin{table}[htbp]
  \centering
  \captionsetup{justification=centering}
  \caption{Pressure Analysis of Microfluidic Units}
  \label{tab:pressure_distribution}
  \begin{tabular}{lcccccccc}
      \hline 
      Length & $P_{in}$ & $P_{out}$ & $\Delta P$ & $\Delta P/L$ & $\text{Velocity}_{in}$ & Flow Rate Q\\
      ($\mu$m) & (Pa) & (Pa) & (Pa) & (Pa/$\mu$m) & (m/s) & ($\mu$L/min)\\
      \hline
      240 & $\approx 850$ & 0 & 850 & 3.54 & 0.0002 & 0.012\\
      240 & $\approx 675$ & 0 & 675 & 2.81 & 0.0005 & 0.03\\
      240 & $\approx 1350$ & 0 & 1350 & 5.63 & 0.001 & 0.06\\
      \hline
  \end{tabular}
 \end{table}

\subsection{Fabrication and Testing}

\begin{figure}[!h]
  \centering
  \includegraphics[width=1\textwidth]{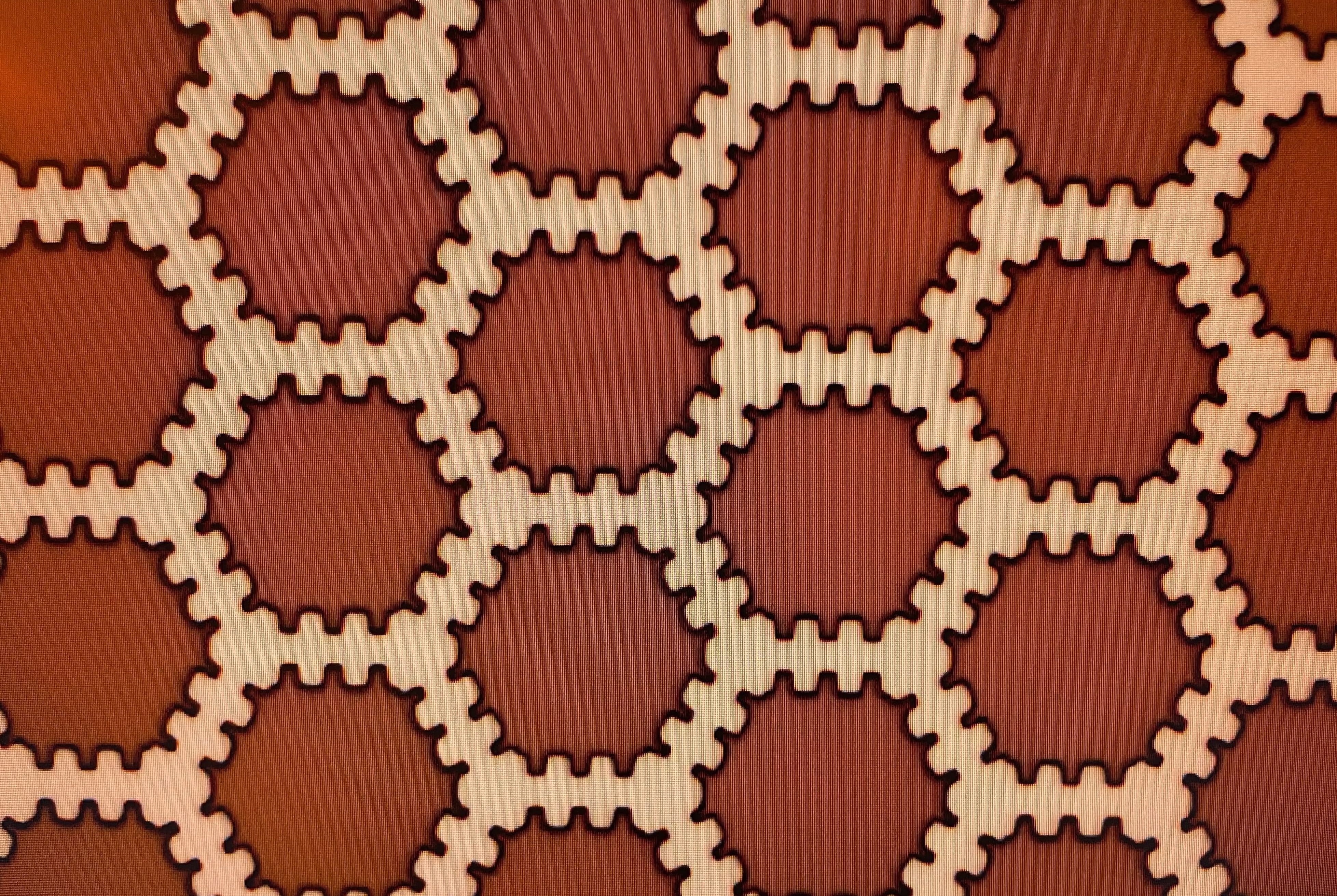}
  \caption{{\bf Microscopy of Fabricated Pattern.}
  Optical microscope image showing the hexagonal-with-gear-teeth array pattern after fabrication.}
  \label{fig:device}
\end{figure}

The photolithographic fabrication process successfully produced the designed hexagonal well patterns. Two distinct pattern types were achieved: smooth-walled hexagonal wells and hexagons with gear-like teeth as shown in Fig.~\ref{fig:device}.

However, examination of the features under microscopy revealed deviations from the intended dimensions. Edge roughness and corner rounding were observed, particularly at the gear-tooth interfaces, possibly indicating that the fabrication resolution approached the limits of the photolithography process. While these imperfections are unlikely to impact the overall functionality of the device, they suggest that further optimization of exposure parameters and development conditions could improve pattern resolution.

While this testing setup was designed and the device was prepared for characterization, experimental flow measurements were not completed in this study. Future work will involve flow characterization and comparison with the computational fluid dynamics predictions to validate device performance.

\section{Conclusion}
This work demonstrated both computational modeling and fabrication capabilities for a future microfluidic platform for organoid computing. The computational fluid dynamics analysis achieved reliable convergence and revealed pressure and velocity distributions suitable for maintaining precise flow control. The fabrication process successfully produced hexagonal well arrays with two distinct patterns though, with some edge roughness at feature boundaries, indicating areas for process optimization.

Future work would focus on: (1) experimental validation of computational predictions through flow characterization, (2) optimization of fabrication parameters to improve pattern fidelity at smaller scales, and (3) integration of the platform with neural network components. These steps will enable the development of a fully functional microfluidic system for future organoid computing projects.

\section{Acknowledgments}
This work was supported by Dr. Eric Stach from the Department of Materials Science and Engineering and Dr. Sam Azadi from the Singh Nanotechnology Center at the University of Pennsylvania. Research guidance in design, fabrication, and simulation was provided by Puyuan Liu from the Department of Materials Science and Engineering and Isabella Huang from the Department of Electrical Engineering, both at the University of Pennsylvania.

\bibliographystyle{unsrtnat}
\bibliography{references}

\end{document}